\documentclass[aps,prb,reprint,groupedaddress,superscriptaddress,twocolumn]{revtex4-1}
\usepackage{graphicx,color,bm,amsmath,amssymb,natbib}

\begin{document}

\newcommand{\MgB}{MgB$_2$}

\newcommand{\ins}[1]{{\textcolor{red}{#1}}}
\newcommand{\note}[1]{{\textcolor{blue}{#1}}}
\newcommand{\me}[1]{\textcolor{purple}{#1}}

\title{Field-angle dependent vortex lattice phase diagram in {\MgB}} 

\author{A.~W.~D.~Leishman}
\affiliation{Department of Physics, University of Notre Dame, Notre Dame, Indiana 46556, USA}

\author{A.~Sokolova}
\affiliation{Australian Centre for Neutron Scattering, Australian Nuclear Science and Technology Organization, NSW 2234, Australia}

\author{M. Bleuel}
\affiliation{NIST Center for Neutron Research, National Institute of Standards and Technology, 100 Bureau Drive, Gaithersburg, Maryland 20899-8562, USA} 
\affiliation{Department of Materials Science and Engineering, University of Maryland, College Park, Maryland 20742-2115, USA}

\author{N.~D.~Zhigadlo}
\altaffiliation[Current address: ]{CrystMat Company, CH-8037 Zurich, Switzerland}
\affiliation{Laboratory for Solid State Physics, ETH, CH-8093 Zurich, Switzerland}

\author{M.~R.~Eskildsen}
\email[email: ]{eskildsen@nd.edu}
\affiliation{Department of Physics, University of Notre Dame, Notre Dame, Indiana 46556, USA}

\date{\today}

\begin{abstract}
Using small-angle neutron scattering we have studied the superconducting vortex lattice (VL) phase diagram in {\MgB} as the applied magnetic field is rotated away from the $c$ axis and towards the basal plane.
The field rotation gradually suppresses the intermediate VL phase which exists between end states aligned with two high symmetry directions in the hexagonal basal plane for $\bm{H} \parallel \bm{c}$.
Above a critical angle, the intermediate state disappears, and the previously continuous transition becomes discontinuous.
The evolution towards the discontinuous transition can be parametrized by a vanishing twelvefold anisotropy term in the VL free energy.
\end{abstract}

\maketitle

\section{Introduction}
Vortex matter in type-II superconductors is highly sensitive to the environment provided by the host material.
An example is the vortex lattice (VL) symmetry and orientation, which is governed by anisotropies in the screening current plane perpendicular to the applied field and the associated nonlocal vortex interactions.\cite{Kogan:1981vl,Kogan:1997vm}
Such anisotropies may arise from the Fermi surface or, in non-$s$ wave superconductors, from nodes in the superconducting gap.
A rich VL phase diagram often arises when this anisotropy is incommensurate with an equilateral triangular VL, as seen in Nb with the applied field along the [100] crystalline direction.\cite{Laver:2006bn, Laver:2009bk, Muhlbauer:2009hw}
However, structural transitions between different VL configurations can also arise when the field is applied perpendicular to a sixfold symmetric crystal plane.
In such cases higher-order contributions to the screening current plane anisotropy become relevant, affecting the orientation of the triangular VL relative to the crystalline axes as seen in UPt$_3$\cite{Gannon:2015ct,Avers:2020wx} and {\MgB}.\cite{Cubitt:2003ip,Das:2012cf}

\begin{figure}[b]
    \includegraphics{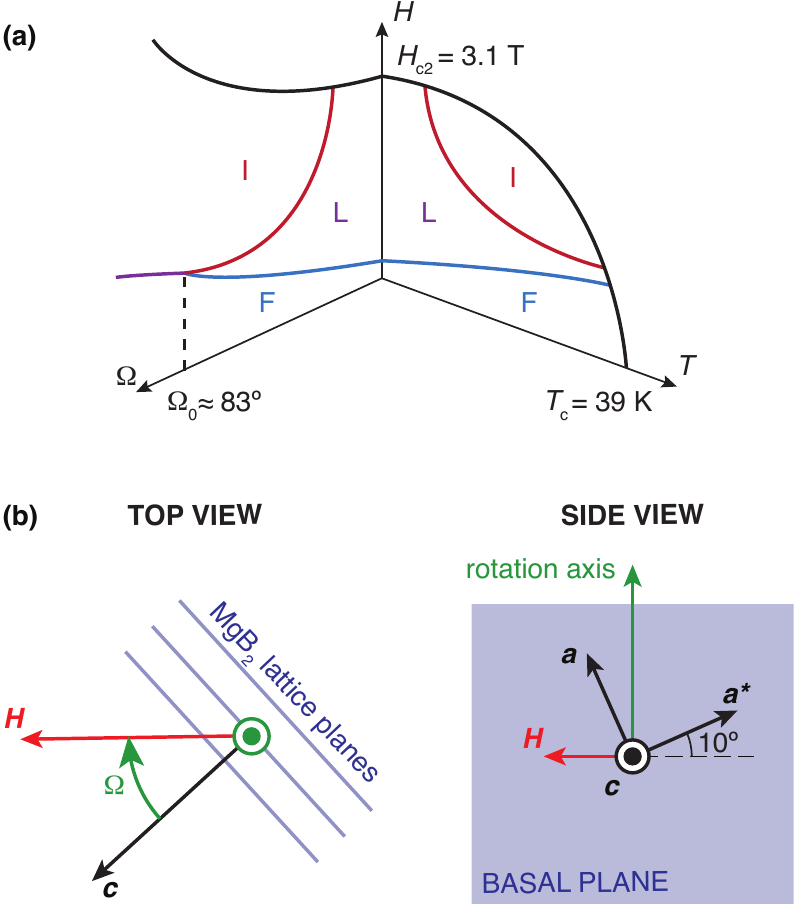}
    \caption{\label{VLPhaseDiagram}
        (a) Qualitative {\MgB} VL phase diagram in the $H\!-\!T$ ($\Omega=0$) and $H\!-\!\Omega$ ($T=0$) planes.
        The former was determined previously,\cite{Das:2012cf} and the latter is discussed in this work.
        The VL L phase vanishes at the critical angle, $\Omega_0$.
        (b) Experimental geometry, indicating the direction and rotation axis of the magnetic field relative to the crystalline axes.
        }
\end{figure}

In the hexagonal superconductor {\MgB}, three different triangular VL phases (labeled F, L and I) were observed for $\bm{H} \parallel \bm{c}$, distinguished by their orientation relative to the crystalline axes.\cite{Das:2012cf}
The multiple phases arise from a competition between the superconducting $\pi$- and $\sigma$-bands which have opposite in-plane anisotropies,\cite{Hirano:2013jx} and are shown in the field-temperature ($H\!-\!T$) plane of Fig.~\ref{VLPhaseDiagram}(a).
For low fields, the $\pi$ band anisotropy causes the VL to align with the $\bm{a}$ crystal axis, producing the F phase.
At high fields, where superconductivity on the $\pi$-band is suppressed,\cite{Eskildsen:2002ih,Cubitt:2003ip} the $\sigma$ band anisotropy aligns the VL with the $\bm{a^*}$ crystal axis, producing the I phase.
At intermediate fields, the VL undergoes a continuous $30^{\circ}$ rotation transition between the two extremes in the L phase.
Since clockwise and counterclockwise rotations are degenerate, the VL fractures into oppositely rotated domains.
This domain formation leads to robust metastable states which inhibit transitions between the three phases and demonstrate activated behavior.\cite{Das:2012cf,Rastovski:2013ff,Louden:2019bq,Louden:2019wx}

Due to the hexagonal crystal structure of {\MgB} and the $s$-wave pairing, the VL free energy for $\bm{H} \parallel \bm{c}$ can be expanded in term with anisotropies that are multiples of six.\cite{Zhitomirsky:2004jq}
Moreover, the continuous rotation in the L phase implies that at least the six- and twelvefold terms are sufficiently strong to influence the VL orientation, as the transition would otherwise be discontinuous.
Here, we have sought to explore the evolution of the VL phase diagram as the twelvefold anisotropy is suppressed by 
rotating the applied field away from the $c$ axis.
We find that the twelvefold anisotropy decreases linearly as the rotation angle is increased, reducing the size of the L phase until it disappears entirely from the phase diagram at a critical value.
Above the critical angle, the VL undergoes a first order phase transition directly from the F to I phase.

\section{Experimental Details}
The VL was studied using small-angle neutron scattering (SANS)~\cite{Muhlbauer:2019jt}.
The SANS measurements were performed at the Bilby instrument\cite{Sokolova:2019jd} at the Australian Nuclear Science and Technology Organization.
Preliminary SANS measurements were carried out at the NG7 beam line\cite{Glinka:1998gd} at the National Institute of Standards and Technology Center for Neutron Research.

The experimental geometry used for the SANS measurements is shown in Fig.~\ref{VLPhaseDiagram}(b).
Here the magnetic field is applied along the horizontal neutron beam direction, and at an angle $\Omega$ relative to the crystalline $c$ axis achieved by rotating the sample about the vertical axis \textit{in situ}.
The SANS measurements used a neutron wavelength $\lambda_n = 0.6$~nm and bandwidth $\Delta \lambda_n/\lambda_n = 10\%$.
All measurements were performed at 2 K.

Measurements were performed on a 200~$\mu$g single crystal of {\MgB} grown using a high pressure cubic anvil technique.\cite{Karpinski:2003aa}
The crystal has a platelet geometry, roughly 1~mm $\times$ 1~mm wide and 50~$\mu$m thick along the $c$ axis, and is isotopically enriched with $^{11}$B to reduce neutron absorption.
The superconducting critical temperature of the sample is $T_{\text{c}} \approx 38$~K, and the upper critical field increases from $H_{\text{c2}} = 3.1$~T to $\sim \! 18$~T as the field is rotated from the $c$ axis ($\Omega = 0$) to the basal plane ($\Omega = 90^{\circ}$).\cite{Karpinski:2003aa}

\section{Results}
\begin{figure}
    \includegraphics[width=\columnwidth]{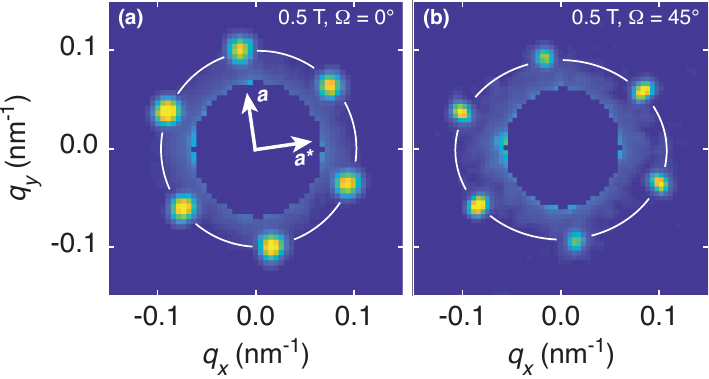}
    \caption{\label{Fig2}
        Vortex lattice diffraction patterns at $\mu_0H = 0.5$~T and (a) $\bm{H} \parallel \bm{c}$ and (b) $\Omega = 45^{\circ}$.
        The orientation of the crystalline axes is indicated in (a).
        Both measurements were performed following a field cooling from above $T_{\text{c}}$.
        Background scattering near the detector center is masked off.
        }
\end{figure}

Figure~\ref{Fig2}(a) shows a VL diffraction pattern obtained for $\bm{H} \parallel \bm{c}$, with all six Bragg peaks lying on a circle of radius $q = 2\pi (2\mu_0 H/\sqrt{3} \Phi_0)^{1/2}$ where $\Phi_0 = h/2e = 2069$~T\,nm$^2$ is the flux quantum.
Here, the field and crystal have been rotated together through the Bragg condition of all six VL peaks.
The crystal was deliberately mounted with the $\bm{a}$ axis roughly 10$^\circ$ from the vertical to investigate whether the degeneracy of the two VL domain orientations in the L phase can be lifted by rotating the applied field away from the $c$ axis.
The field rotation introduces a distortion of the VL due to the different penetration depth within the basal plane vs perpendicular to it.\cite{Campbell:1988vf}
This is seen in the SANS diffraction patterns in Fig.~\ref{Fig2}(b) as a relocation of the six Bragg peaks such that they lie on an ellipse with the minor axis parallel to the axis of rotation.
We note that the diffraction patterns in Fig.~\ref{Fig2} were obtained following a field cooling, which for $\bm{H} \parallel \bm{c}$ left the VL in a metastable F phase.\cite{Das:2012cf}

\begin{figure}
    \includegraphics[width=\columnwidth]{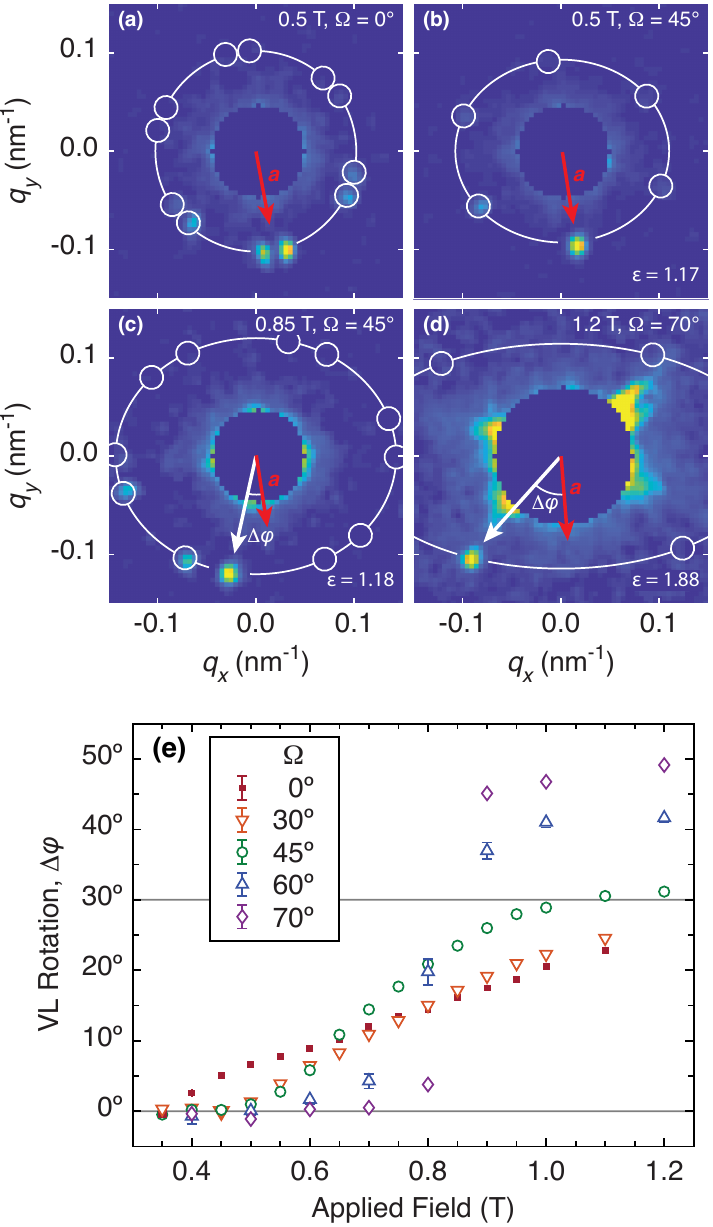}
    \caption{\label{DiffPat}
        Vortex lattice diffraction patterns at (a) $H = 0.5$~T and $\Omega = 0^{\circ}$, (b) 0.5~T and $45^{\circ}$, (c) 0.85~T and $45^{\circ}$, and (d) 1.2~T and $70^{\circ}$.
        In all cases only a single (in (a) and (c) split) peak fully satisfies the Bragg condition.
        The other peaks are indicated by open circles in their predicted locations.
        For each diffraction pattern the fitted geometric anisotropy ($\varepsilon$) is indicated.
        All measurements were performed following a damped field oscillation.
        Background scattering near the detector center is masked off.
        (e) Rotation of the VL Bragg peaks around $q = 0$ as a function of field for different $\Omega$.
        Angles are measured with respect to the VL orientation at 0.3~T (F phase).
        Error bars represent one standard deviation. 
        }
\end{figure}

To test the effect of the field rotation on the VL phase diagram, measurements were made with $\Omega = 0^{\circ}, 30^{\circ}, 45^{\circ}, 60^{\circ}$, and $70^{\circ}$ and fields between 0.3~T and 1.2~T.
Prior to each measurement, the magnitude of the applied field was cycled through a damped oscillation about the desired value with large enough amplitude (+5 mT, -4 mT, +3 mT, -2 mT, +1 mT) to overcome the activated behavior in {\MgB} and ensure an equilibrium VL configuration.
For $\Omega = 0$, this produces an equilibrium L phase configuration at 0.5~T shown in Fig.~\ref{DiffPat}(a) as opposed to the metastable F phase configuration shown in Fig.~\ref{Fig2}(a).\cite{Levett:2002ba,Das:2012cf, Louden:2019bq}
As seen in Figs.~\ref{DiffPat}(a)-\ref{DiffPat}(d), the elliptical distortion increases with $\Omega$.
It is useful to quantify this distortion by the geometric anisotropy of the ellipse $\varepsilon$, defined as the ratio of its major and minor axes.
At $\Omega = 90^{\circ}$, $\varepsilon$ is expected to reach the penetration depth anisotropy~\cite{Campbell:1988vf}.
To conserve beam time not all VL peaks were rocked through the Bragg condition for all measurements; however, their location can be determined from symmetry and the analysis discussed below, and are indicated by open circles in Figs.~\ref{DiffPat}(a)-\ref{DiffPat}(d).
Within the L phase, split Bragg peaks corresponding to the two degenerate VL orientations are observed as seen in Figs.~\ref{DiffPat}(a) and \ref{DiffPat}(c).

The angular rotation $\Delta \varphi$ of the VL Bragg peak as a function of field and $\Omega$ is summarized in Fig.~\ref{DiffPat}(e).
At each $\Omega$ the rotation is measured relative to the peak position in the F phase at 0.3~T, corresponding to the projection of the $\bm{a}$ axis onto the scattering plane, as shown in Figs.~\ref{DiffPat}(c) and \ref{DiffPat}(d) and given by $\tan \varphi' = \tan \varphi \cdot \cos \Omega$.
For $\Omega = 0$, we observe the same VL rotation reported earlier,\cite{Das:2012cf} associated with the progression through the L phase.
In the L phase, where the VL Bragg peaks are split, we show the orientation of the clockwise rotating domain.
As $\Omega$ is increased, the VL distortion allows the rotation of peaks which start near the minor axis to exceed the $30^{\circ}$ range for the F to I transition when $\bm{H} \parallel \bm{c}$.
This is seen most clearly at $\Omega = 70^{\circ}$, where the VL rotation approaches $50^{\circ}$ at 1.2~T.

\begin{figure*}
    \includegraphics{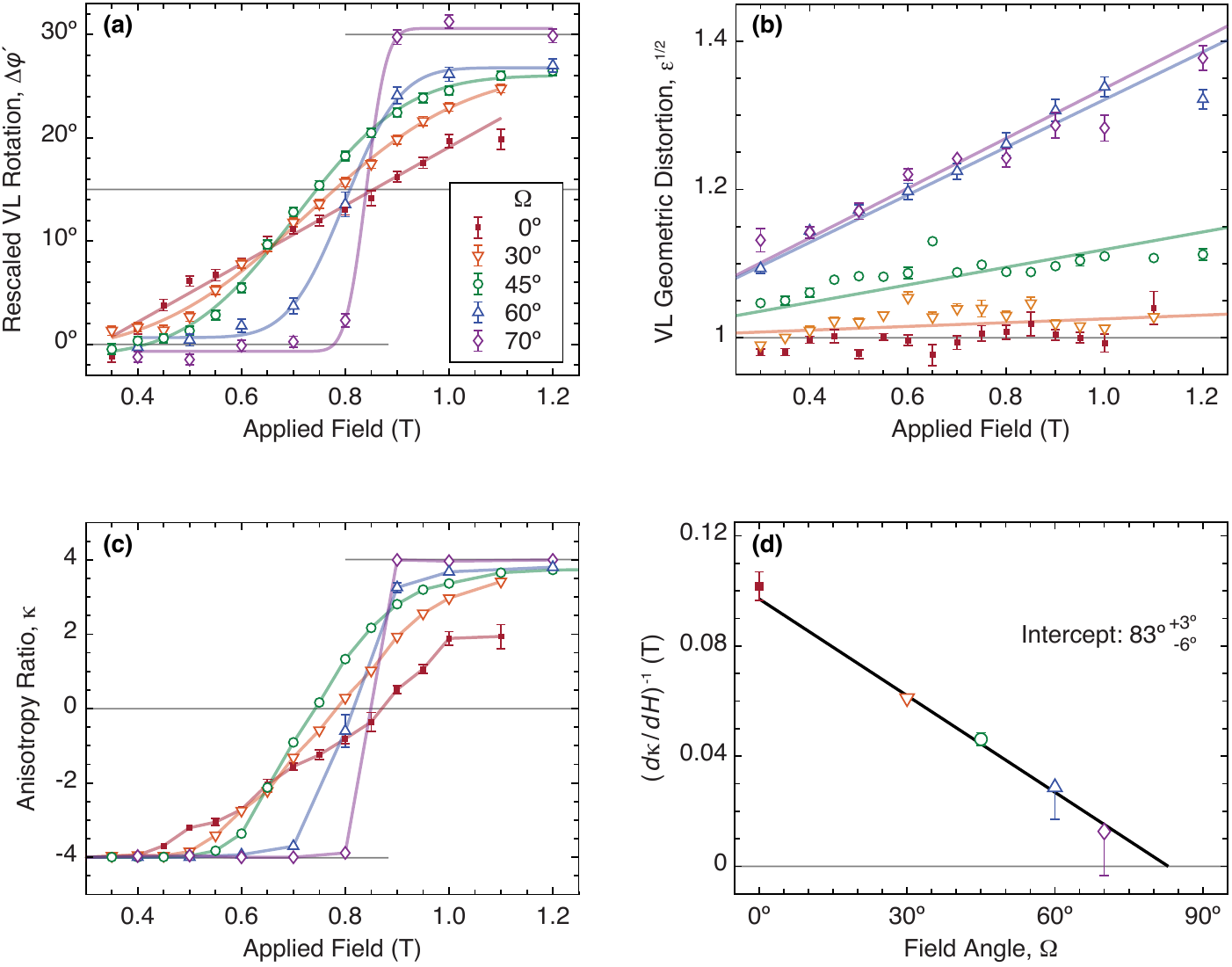}
    \caption{\label{GammaFig}
        Evolution of VL rotation transition versus applied field for increasing field angle:
        (a) rescaled VL rotation corresponding to Fig.~\ref{DiffPat}(e) using Eqs.~(\ref{gamma}) -- (\ref{DeltaPhi}),
        (b) square root of the geometric distortion, $\varepsilon$, including linear fits, and
        (c) anisotropy ratio, $\kappa$,
        (d) inverse slope of the anisotropy ratio near $\kappa$ = 0 as a function of $\Omega$.
        In all panels, error bars represent one standard deviation.}
\end{figure*}

The location of the VL peak positions in the SANS data are governed by two separate effects: the VL rotation transition within the L phase, and the geometric distortion due to the penetration depth anisotropy discussed above.
To analyze the progression of the rotation transition it is useful to first remove the effect of the geometric distortion,  which corresponds to mapping the VL Bragg peaks positions from lying on an ellipse back onto a circle.
Due to flux quantization, the area of the circle in reciprocal space must be the same as the original ellipse, and the transformation can treated as a squeeze mapping of all points $(q_x',q_y')$ in the circle to all points $(q_x,q_y)$ in the ellipse: 
\begin{equation}
    \label{squeeze}
    \left( \begin{array}{c} q_x' \\ q_y' \end{array} \right)
      = \left( \begin{array}{cc} \varepsilon^{-1/2} & 0 \\ 0 & \varepsilon^{1/2} \end{array} \right)
        \left( \begin{array}{c} q_x \\ q_y \end{array} \right).
\end{equation}
Converting to polar coordinates $(q',\varphi')$ and $(q,\varphi)$, and solving for $\varepsilon$ yields a transcendental equation that can be solved numerically given any point on the ellipse:
\begin{equation}
  \label{gamma}
  \varepsilon^{1/2} = \frac{q'}{q\sin{\varphi}} \sin \left( \arccos \left[ \frac{q \, \cos{\varphi}}{q' \, \varepsilon^{1/2}} \right] \right).
\end{equation}
Values of $q$ and $\varphi$ are determined from 2D Gaussian fits to VL Bragg peaks on the SANS detector, and $q'$ is determined from the measurement performed at the same applied field with $\Omega$ = 0.
Once $\varepsilon$ is determined, the re-scaled VL orientation $\varphi'$ is given by
\begin{equation}
    \label{theta}
    \varphi' = \arccos \left[\frac{q \, \cos{\varphi}}{q' \, \varepsilon^{1/2}} \right].
\end{equation}
Finally, the rescaled VL rotations can be calculated by subtracting the VL orientation at 0.3~T,
\begin{equation}
    \label{DeltaPhi}
    \Delta\varphi'(H,\Omega) = \varphi'(H,\Omega) -\varphi'(0.3~T,\Omega),
\end{equation}
and are shown in Fig.~\ref{GammaFig}(a).
The transformation rescales the VL rotation so that the phase transition always occurs in the range $\Delta\varphi' =  0^\circ$ to $30^\circ$ for all $\Omega$, consistent with the L phase behavior for $\bm{H} \parallel \bm{c}$.

From Fig.~\ref{GammaFig}(a) it is clear that the VL phase diagram changes dramatically with increasing $\Omega$.
Firstly, the onset of the rotation transition moves to higher fields, indicating that the F phase ($\Delta \varphi' = 0^{\circ}$) is expanding.
Secondly, once the rotation does begin, it occurs more rapidly at high $\Omega$.
Thirdly, the L phase ($0^{\circ} < \Delta \varphi'<30^{\circ}$)  either becomes very narrow or vanishes entirely for $\Omega > 70^{\circ}$.
This would correspond to a discontinuous phase transition directly from the F to the I phase ($\Delta \varphi' = 30^{\circ}$).
The increasingly abrupt rotation transition is also evident in Figs.~\ref{GammaFig}(b), which shows an abrupt jump in the slope of $\sqrt{\varepsilon}$ versus field for $\Omega \geq 60^{\circ}$.
Here, we note that the field dependence of $\varepsilon$ at {\em constant} $\Omega$ is due to the gradual suppression of superconductivity on the $\pi$-band.\cite{Cubitt:2003ip}
This rise in $\varepsilon$ with respect to field is similar to phenomena observed in V$_3$Si,\cite{Christen:1985wo} 2$H$-NbSe$_2$,\cite{Gammel:1994vn} and Sr$_2$RuO$_4$.\cite{Kuhn:2017bv}

\section{Discussion}
The VL phase diagram in {\MgB} can be modeled by a free energy
\begin{equation}
    \label{FreeEnergy}
    \delta F(\Delta \varphi') = K_6\cos{(6 \, \Delta \varphi')} + K_{12}\cos{(12 \, \Delta \varphi')},
\end{equation}
containing six- and twelve-fold anisotropy terms.
This form of the free energy was originally proposed for $\bm{H} \parallel \bm{c}$ where $\varphi' = \varphi$,\cite{Zhitomirsky:2004jq,Das:2012cf,Olszewski:2020jy} but here it is generalized to non-zero $\Omega$.
In this model $K_{12}$ is positive, and the continuous rotation transition in the L phase occurs when the anisotropy ratio  $\kappa = K_6/K_{12}$ is varied between $-4$ and 4.
Note that if $K_{12}$ was negative, the VL would undergo a discontinuous reorientation transition between the F and I phases when $K_6$ changes sign.
For each measurement it is possible to calculate the anisotropy ratio
\begin{equation}
    \label{kappa}
    \kappa = -4 \cos{(6 \, \Delta \varphi')},
\end{equation}
obtained from a minimization of the free energy in Eq.~(\ref{FreeEnergy}).
This is shown in Fig.~\ref{GammaFig}(c), and is seen to mimic the behavior of the re-scaled rotation angle $\Delta \varphi'$.
We note that actual values of $\kappa$ most likely extend outside the $\pm 4$ range allowed by Eq.~(\ref{kappa}).
However, this does not affect the analysis below as it focuses on the region where $\kappa$ is close to zero.

The increasingly abrupt transition seen in Figs.~\ref{GammaFig}(a) and \ref{GammaFig}(c) suggests a reduction and possible sign change of $K_{12}$ as $\Omega$ increases.
It is possible to model the increasingly rapid reorientation transition by Taylor expanding $K_6$  to linear order around the field $H_0$ where $K_6$, and thus $\kappa$, vanish:
\begin{equation}
    \label{TEk6}
    K_{6}(H) \approx \alpha_6(H - H_0).
\end{equation}
Here, $\alpha_6$ is an undetermined constant.
The slope of $\kappa$ with respect to field near the transition at $H = H_0$ is then an indirect measure of $K_{12}$:
\begin{equation}
    \left. \frac{\partial \kappa}{\partial H} \right|_{H = H_0} = \frac{\alpha_6}{K_{12}},
\end{equation}
where we have taken $K_{12}$ to be field independent.
Figure~\ref{GammaFig}(d) shows the inverse slope $(\partial \kappa/\partial H)^{-1}$ measured near $\kappa = 0$ for each $\Omega$. The inverse slope depends linearly on $\Omega$, suggesting a functional form for $K_{12}$:
\begin{equation}
    \label{TEk12}
    K_{12}(\Omega) = \alpha_{12}(\Omega - \Omega_0).
\end{equation}
Extrapolating to $K_{12}/{\alpha_6} = 0$, we find an estimate for $\Omega_0 = \left( 83^{+3}_{-6}\right)^{\circ}$, where the error bars both represent one standard deviation.
For values of $\Omega > \Omega_0$, $K_{12}$ will be negative and the VL will undergo a first-order transition between the F and I phases.
As a result, the L phase vanishes and $H_0$ becomes a critical field corresponding to an equal F and I phase energy (rather than simply the field where $\Delta \varphi' = 15^{\circ}$ which is the case for $\Omega < \Omega_0$).
We note that $\Omega = 70^{\circ}$ may already be above $\Omega_0$, as the uncertainly on $(\partial \kappa/\partial H)^{-1}$ at this angle extends to negative values.
Behavior qualitatively similar to {\MgB} has also been observed in YBCO,\cite{Yethiraj:1993iv} where a triangular to square VL transition switches from second to first order at a critical angle of $10^{\circ}$. The much smaller $\Omega_0$ for YBCO is likely due to the highly two-dimensional nature of this material, making it relatively more susceptible to field rotations away from the $c$ axis.

Finally, we return to the question of whether rotating the applied field away from the $c$ axis will break the VL domain degeneracy within the L phase.
Such an effect was previously observed in TmNi$_2$B$_2$C where the VL undergoes a triangular to square transition with a degenerate intermediate rhombic phase,\cite{Eskildsen:1998aa} and where rotating the field away from the $c$ axis by $\approx 10^{\circ}$ is sufficient to suppress one of the two rhombic domains.\cite{DewhurstPC}
In an ideal uniaxial superconductor with an isotropic basal plane, London theory predicts that the two-fold anisotropy introduced by field rotation will favor a VL orientation with Bragg peaks on the minor axis of the ellipse in reciprocal space.\cite{Campbell:1988vf}
However, any real material will exhibit some basal plane anisotropy which may compete with the uniaxial effect, and the relative strength of the two will determine the VL orientation.
As an example, one can consider NbSe$_2$, where the triangular VL is oriented in a manner corresponding to the maximal energy according to the London theory.\cite{Gammel:1994vn}

In our SANS measurements the {\MgB} crystal was mounted such that within the L phase Bragg peaks for one of the domains lie near the vertical axis, as seen in Fig.~\ref{DiffPat}(a), and are thus favored by the uniaxial effect.
However, peaks corresponding to both L phase domains were clearly observed at  $\Omega = 45^{\circ}$and 0.6~T and at $60^{\circ}$ and 0.9~T, indicating that the degeneracy is not readily lifted in {\MgB}.
This suggests that at these values of $\Omega$ the uniaxial anisotropy is weaker than $K_{12}$, and that a suppression of the degeneracy will only occur just below $\Omega_0$.
It is also consistent with the inability of the uniaxial effect to cause a VL reorientation in the F phases to have the Bragg peaks on the minor axis of the ellipse, since $K_6$ and $K_{12}$ are of the same order of magnitude except in the immediate vicinity of $H_0$ and $\Omega_0$.
That said, the uniaxial anisotropy does appear to lower the free energy of the F phase relative to the I phase, causing $H_0$ to shift to higher fields in Figs.~\ref{GammaFig}(a) and \ref{GammaFig}(c) as $\Omega \rightarrow \Omega_0$.
This conclusion is supported by previous SANS measurements where the magnetic field was rotated about the $\bm{a}^*$.\cite{Pal:2006aa}
Here the L phase (rather than the F phase) is favored by the uniaxial anisotropy, and at 0.4~T the critical angle is shifted to a lower value $\Omega_0 \sim 70^{\circ}$.

\section{CONCLUSION}
\label{Sec:Conclusion}
In summary, we have demonstrated how the continuous (second order) VL rotation transition observed in {\MgB} for $\bm{H} \parallel \bm{c}$ gradually evolves towards and finally becomes discontinuous (first order) as the applied magnetic field is rotated away from the $c$ axis by $\approx 83^{\circ}$.
For rotation angles below this critical value,  domain formation in the intermediate L phase persists.
We speculate that the long lived metastable VL phases, attributed to domain formation in the L phase, will thus no longer occur above the critical angle, and should be the subject of further SANS studies.

\section*{Acknowledgements}
This work was supported in part by the U.S. Department of Energy, Office of Basic Energy Sciences, under Award No. DE-SC0005051 (A.W.D.L, M.R.E). We acknowledge the support of the National Institute of Standards and Technology, U.S. Department of Commerce, in providing the neutron research facilities used in this work.
We acknowledge useful conversations with G.~Longbons and A.~Francisco in preparation for these experiments, as well as the assistance of J.~Krzywon, A.~Qiang, and C.~Baldwin in completing them.

\bibliography{MgB2anisoFinal}

\end{document}